\documentclass[conference]{IEEEtran}
\IEEEoverridecommandlockouts
\usepackage{cite}
\usepackage{amsmath,amssymb,amsfonts,amsthm}
\usepackage{algorithm,algorithmic}
\usepackage{graphicx}
\usepackage{textcomp}
\usepackage{xcolor}
\newtheorem{proposition}{Proposition}

\def\BibTeX{{\rm B\kern-.05em{\sc i\kern-.025em b}\kern-.08em
    T\kern-.1667em\lower.7ex\hbox{E}\kern-.125emX}}
\begin{document}

\title{Millimeter Wave Channel Estimation for Lens based  Hybrid MIMO with Low Resolution ADCs}

\author{\IEEEauthorblockN{Evangelos Vlachos$^1$, Aryan Kaushik$^2$ and Muhammad Z. Shakir$^3$} 
\IEEEauthorblockA{
$^1$Industrial Systems Institute, Athena Research and Innovation Centre, Greece.\\
$^2$School of Engineering and Informatics, University of Sussex, United Kingdom.\\
$^3$School of Computing, Engineering \& Physical Sciences, University of the West of Scotland, United Kingdom. \\
E-mails: evlachos@athenarc.gr, aryan.kaushik@sussex.ac.uk, muhammad.shakir@uws.ac.uk}} 
\maketitle

\begin{abstract}
The high path loss associated with millimeter wave (mmWave) frequency communication can be compensated by large scale antenna arrays such as multiple-input multiple-output (MIMO) systems. The hybrid beamforming architecture which uses fewer radio frequency chains is implemented to reduce power consumption and hardware complexity, while still supporting multi-stream communication. We propose an efficient expectation-maximization (EM)-based mmWave channel estimator for a lens-based hybrid MIMO system with low resolution sampling at the receiver. The lens-based beamformer is investigated to provide increased antenna gain and reduced implementation complexity as the conventional beam selection network is excluded. Low resolution sampling at the analog-to-digital converters is implemented for reduced power consumption. The proposed solution with a robust maximum a posteriori estimator based on the EM algorithm performs better than the conventional EM approach and minimum mean square error baselines in medium to high signal-to-noise ratio regions.
\end{abstract}


\section{Introduction}
The next generation standards require increased capacity, high data rates, and an availability of accurate channel state information at transceivers. The fifth generation (5G) speeds are being forecast 13 times higher than average mobile connection by 2023 \cite{cisco2020}, and 5G subscriptions being 440 million just in Western Europe by 2027 \cite{ericsson2022}. The microwave frequency spectrum at sub-6 GHz frequencies, which we currently use for mobile broadband, is limited to a very crowded frequency range. This increases the demand for unused but available spectrum which can be resolved by the use of millimeter wave (mmWave) frequency spectrum \cite{rangan2014}. Large-scale antenna arrays such as multiple-input multiple-output (MIMO) systems can be incorporated with mmWave communication to compensate for high path loss at such high frequency. 

The high complexity of mmWave MIMO systems can be significantly reduced through hybrid beamforming (HBF) where the number of radio frequency (RF) chains and associated analog-to-digital converters (ADCs) and digital-to-analog converters (DACs) are much less than the number of antennas \cite{ahmadiTWC2009}. 
Such systems can also be optimized to achieve high energy efficiency (EE) \cite{aryanTGCN2019, aryanJSAC2019, vlachosRsoc2020}. More recent applications of low resolution DACs, RF optimization and low complexity architectures has been into the next generation joint radar-communication systems \cite{aryanICC2021, aryanICC2022, dizdarOJCOMS2022, aryanJCNS2022, aryanIET2022}.

In addition to the phase-shifter-based HBF systems, lens-based beamforming systems can be implemented as a practical alternative to the conventional beamforming architectures \cite{constk}. These approaches simplify the mmWave MIMO RF front-end, such as Rotman lens based MIMO system with beam selection and digital beamforming is discussed in \cite{7946172}. Reference \cite{8693726} discusses a broadband mmWave analog beamforming design based on Rotman lens antenna array, whereas a practical two-stage Rotman lens analog beamformer is shown in \cite{abbasiECAP2019}. A lens antenna array enabled mmWave MIMO communication is discussed in \cite{wangJSAC2017} and HBF is a feasible candidate for mmWave MIMO systems to reduce power consumption and hardware complexity. 
Implementing low resolution quantization in mmWave MIMO systems can reduce power consumption and complexity of such systems \cite{aryanICC2019}. 

Estimating the channel in mmWave MIMO systems is a challenging problem due to the distinct channel characteristics of the sparse mmWave channel which experiences less multipath reflection and refraction than at conventional sub 6-GHz microwave frequencies. Advanced channel estimation approaches have attracted attention in recent years such as for the case of reconfigurable intelligent surfaces (RIS) and unmanned aerial vehicles (UAVs)-aided communication \cite{alexandropoulos2020, alexandropoulos2020_2}. Reference \cite{aryanEUSIPCO2018} discuss sparse channel estimation solutions for mmWave MIMO systems where low resolution ADCs are employed at the RX. References \cite{gaoTWC2017, yangTVT2018} discuss channel estimation for mmWave MIMO systems with lens antenna arrays where they exploit mmWave sparsity, and \cite{gaoWCSP2016} discusses channel estimation scheme for a three dimensional beamspace channel model in a lens-based mmWave massive MIMO system. However, channel estimation for lens enabled mmWave hybrid MIMO systems is not widely studied in existing literature and low resolution sampling has not been exploited for such systems. 

\subsubsection*{Contributions} This paper proposes an efficient channel estimator for Rotman lens-based mmWave hybrid MIMO system with low resolution ADCs at the RX. We exploit the sparsity of the mmWave channel and use the beamspace representation to formulate sparse signal recovery problem. We then implement a robust maximum a posteriori (MAP) estimator based on the expectation maximization (EM) algorithm. 
The simulation results verify that the proposed robust EM approach performs better than the conventional EM approach and minimum mean square error (MMSE) baselines in terms of estimation accuracy, i.e., mean square error (MSE), and exhibits low computational complexity.


\emph{Notation:} $\mathbf{A}$, $\mathbf{a}$, and $\textit{a}$ denote a matrix, a vector, and a scalar, respectively. The complex conjugate transpose, transpose and complex conjugate of matrix $\mathbf{A}$ are denoted as $\mathbf{A}^{H}$, $\mathbf{A}^{T}$ and $\mathbf{A}^*$, respectively; $\textrm{vec}(\mathbf{A})$ denotes the vector of entries of the matrix $\mathbf{A}$; $\mathbf{I}_\textrm{N}$ represents $N \times N$ identity matrix; $\mathbf{X} \in \mathbb{C}^{A \times B}$ and $\mathbf{X} \in \mathbb{R}^{A \times B}$ denote an $A \times B$ size $\mathbf{X}$ matrix with complex and real entries, respectively; $\mathbf{X}\otimes\mathbf{Y}$ denotes the Khatri-Rao product of the $\mathbf{X}$ and $\mathbf{Y}$ matrices; $\mathcal{C}\mathcal{N} (\mathbf{a}; \mathbf{A})$ denotes a complex Gaussian vector having mean $\mathbf{a}$ and covariance matrix $\mathbf{A}$; the expectation of a complex variable is denoted as $\mathcal{E} (\cdot)$; $[\mathbf{A}]_k$ denotes the $k$-th column of matrix $\mathbf{A}$ and $[\mathbf{A}]_{kl}$ is the matrix entry at the $k$-th row and $l$-th column.


\section{Lens-based MmWave Hybrid MIMO Model}

\subsection{MmWave Channel Model}
Let us denote the antenna array size at the RX as $M$ and we assume the TX has only a single antenna. 
We model the $M \times 1$ uplink channel vector $\mathbf{h}_{\ell}$ for the $\ell$-th terminal as a double-directional response, consisting of a finite number of multipath components $N_{\textrm{P}}$, as follows: 
\begin{equation}
\label{propagationchanneltoUEl}
\mathbf{h}_{\ell}=\frac{1}{\sqrt{N_{\textrm{P}}}}
\sum\limits_{p=1}^{N_{\textrm{P}}}\hspace{1pt}
\alpha_{\ell,p}\hspace{2pt}
\Lambda\left(\phi_{\ell,p},\theta_{\ell,p}\right)
\mathbf{a}^{H}\hspace{-2pt}
\left(
\phi_{\ell,p},\theta_{\ell,p}\right), 
\end{equation}
where $\alpha_{\ell,p}$, $\Lambda\left(\phi_{\ell,p},\theta_{\ell,p}\right)$ and $\mathbf{a}\left(\phi_{\ell,p},\theta_{\ell,p}\right)$ denote the gain 
of the $p$-th multi-path component, the per-antenna element 
gain and the far-field steering vector of the uniform rectangular array (URA) \cite{8108586}, respectively. Also, $\alpha_{\ell,p}\sim
\mathcal{CN}\left(0,\beta_{\ell}\right)$ when 
$\beta_{\ell}=\zeta_{\ell}
(r_{\textrm{ref}}/r_{\ell})^{\chi}$ captures the large-scale fading impact within the channel, involving the shadow fading and geometric attenuation with the distance $r_{\ell}$ from the $\ell$-th user equipment (UE) to the URA. In particular, 
$10\log_{10}(\zeta_{\ell})\sim\mathcal{CN}
\left(0,\sigma_{\textrm{sf}}^{2}\right)$, where 
$\sigma_{\textrm{sf}}$ is the standard deviation of the shadow fading. Here $r_{\textrm{ref}}$ is the 
reference distance from URA, while $\chi$ is the attenuation 
exponent. 

For mmWave channel estimation, we require the beamspace representation \cite{brady, dai} of the channel in \eqref{propagationchanneltoUEl} which can be expressed as follows:
\begin{equation}
\mathbf{H} = \mathbf{D}_\textrm{R} \mathbf{Z} \mathbf{D}_\textrm{T}^{H},
\end{equation}
where $\mathbf{Z} \in \mathbb{C}^{N_\textrm{R} \times N_\textrm{T}}$ represents a sparse channel matrix with a few non-zero entries which are assumed to follow a Bernoulli-Gaussian distribution, while $\mathbf{D}_\textrm{R} \in \mathbb{C}^{N_\textrm{R} \times N_\textrm{R}}$ and $\mathbf{D}_\textrm{T} \in \mathbb{C}^{N_\textrm{T} \times N_\textrm{T}}$ are the discrete Fourier transform (DFT) matrices. By following the basic matrix multiplication property, i.e., $\textrm{vec}(\mathbf{A}\mathbf{X}\mathbf{B}) = (\mathbf{B}^T\otimes\mathbf{A}) \textrm{vec}(\mathbf{X})$, we can express the following:
\begin{equation}\label{eq:beamspace}
\textrm{vec}(\mathbf{H}) = \textrm{vec}(\mathbf{D}_\textrm{R} \mathbf{Z} \mathbf{D}_\textrm{T}^{H}) = (\mathbf{D}_\textrm{T}^{*}\otimes \mathbf{D}_\textrm{R}) \textrm{vec}(\mathbf{Z}),    
\end{equation}
where $\mathbf{z} = \textrm{vec}(\mathbf{Z})$, i.e., the mmWave channel vector to be estimated in Section III below.

\subsection{Quantization Model}

We consider the outputs of the two dimensional Rotman lens-based beamformer to be connected to $L$ parallel RF chains which represent the received complex signal vector $\mathbf{y} \in \mathbb{C}^{L \times 1}$. This signal passes through $L$ ADCs with low-resolution quantization, i.e., $b \in \{3, 4, 5\}$ bits for each real or imaginary component. Hence,
\begin{equation}
    \mathbf{r} = \mathcal{Q}\big(\mathtt{Re}(\mathbf{y})\big) + j \mathcal{Q}\big(\mathtt{Im}(\mathbf{y})\big)
\end{equation}
where $\mathcal{Q}(\cdot)$ represents a uniform symmetric mid-riser type quantizer which is applied independently at each component of the input vector. Specifically, for the scalar input $x$ it is defined as:
\begin{equation}
    \mathcal{Q}(x) = \textrm{sign}(x) \left[ \min \left( \left\lceil \frac{\vert x \vert}{\delta} \right\rceil, 2^{b-1} \right) - \frac{1}{2} \right] \delta
\end{equation}
where $\delta \triangleq \big( \mathcal{E}(\vert x \vert^2) \big)^{1/2} \gamma$  and $\gamma$ is the quantization stepsize. We assume that the average power $\mathcal{E}(\vert x \vert^2)$ is known, measured by an automatic gain control unit, while the stepsize $\gamma$ is chosen optimally so as to minimize the quantization error assuming a Gaussian input signal.

\subsection{Lens-based Hybrid MIMO System Model}
We consider a base station (BS) equipped with an URA followed by a Rotman lens-based hybrid MIMO beamformer. In a traditional MIMO system at sub 6-GHz microwave frequencies, each antenna element is connected to one RF chain which is followed by the digital baseband processing unit. In mmWave MIMO systems, we make use of HBF where there are fewer RF chains than the number of antennas. We implement the HBF architecture where we can use a smaller number of RF chains with a larger number of antenna elements, e.g., 15 (3 Rotman lenses with access points $N_{\textrm{ap}} = 5$). 

The $L \times 1$ down-converted signal $\mathbf{y} \in \mathbb{C}^{L \times 1}$ can be written as
\vspace{-6pt}
\begin{equation}
\label{receivedsignal}
\mathbf{y}=\rho^{1/2}\mathbf{F}_{\textrm{RF}}
\mathbf{H} \mathbf{s} +\mathbf{n},
\end{equation}
where $\mathbf{s} \in \mathbb{C}^{L \times 1}$ is the $L \times 1$ transmitted symbol vector from all terminals, $\mathbf{H} \in \mathbb{C}^{N_\textrm{R} \times L}$ is the mmWave channel matrix, $\mathbf{F}_\textrm{RF} \in \mathbb{C}^{L \times N_\textrm{R}}$ is the lens-based beamformer matrix and $\mathbf{n} \in \mathbb{C}^{L \times 1}$ is the additive complex Gaussian noise vector with $\mathbf{n} \sim \mathcal{CN}(\mathbf{0},\sigma_\textrm{n}^2 \mathbf{I}_\textrm{L})$. We assume that the beamformer is capable of creating $M$ fixed analog beams along azimuth and elevation sectors as $(\phi_1, \theta_1), \ldots, (\phi_M, \theta_M)$. The net functionality of the lens-based beamformer when considering a perfect focusing capability is described by the 
$L\times{}M$ matrix: 
\vspace{-2pt}
\begin{equation}
\label{idealbeamformer}
\mathbf{F}_{\textrm{RF}}=\left[\hspace{1pt}
\mathbf{a}^{H}\hspace{-1pt}\left(\phi_{1},\theta_{1}\right)\hspace{3pt}
\mathbf{a}^{H}\hspace{-1pt}\left(\phi_{2},\theta_{2}\right)\hspace{3pt}
\dots\hspace{3pt}
\mathbf{a}^{H}\hspace{-1pt}
\left(\phi_{M},\theta_{M}\right)\hspace{1pt}\right]^{T}. 
\vspace{0pt}
\end{equation}

Furthermore, \eqref{receivedsignal} can be re-written, by following the property of $\textrm{vec}(\mathbf{A}\mathbf{X}\mathbf{B}) = (\mathbf{B}^T\otimes\mathbf{A}) \textrm{vec}(\mathbf{X})$, as following:
\begin{equation}
\label{receivedsignal_update}
\mathbf{y}=\rho^{1/2}\mathbf{F}_{\textrm{RF}} (\mathbf{s}^T \otimes \mathbf{I})
\textrm{vec}(\mathbf{H})  +\mathbf{n}.
\end{equation}
Following \eqref{eq:beamspace}, we can express the signal $\mathbf{y}$ in terms of the channel vector $\mathbf{z}$ as follows:
\begin{align}\label{eq:received_beamspace}
\mathbf{y}=& \rho^{1/2}\mathbf{F}_{\textrm{RF}} (\mathbf{s}^T \otimes \mathbf{I})
(\mathbf{D}_\textrm{T}^{*}\otimes \mathbf{D}_\textrm{R}) \textrm{vec}(\mathbf{Z}) +\mathbf{n} \nonumber \\
=& \rho^{1/2}\mathbf{F}_{\textrm{RF}} (\mathbf{s}^T \otimes \mathbf{I})
(\mathbf{D}_\textrm{T}^{*}\otimes \mathbf{D}_\textrm{R}) \mathbf{z} +\mathbf{n}.
\end{align}

\section{Proposed Channel Estimator}

\subsection{Problem Formulation}

We consider a two-stage stacked Rotman lens-based beamformer, which eliminates the requirement of a switching matrix \cite{abbasiECAP2019}, which directly reduces the complexity of a mmWave hybrid MIMO architecture. However, closely spaced antenna elements are subject to mutual coupling that has a negative impact on per antenna element efficiency and there are inherent losses associated with the Rotman lens \cite{ESA}. We consider realistic beam patterns for the Rotman lens system in an anechoic chamber. Thus, we model these imperfections of the antenna and the beamformer according to
\begin{equation}\label{eq:beamformer_losses}
    \mathbf{F}_{\textrm{RF}} = \mathbf{E}_m \circ \mathbf{F}_{\textrm{RF}}^o + \mathbf{E}_a
\end{equation}
where $\mathbf{E}_m, \mathbf{E}_a \in \mathbb{C}^{L \times N_{\textrm{T}}}$ represent the beamformer imperfections and $\mathbf{F}_{\textrm{RF}}^o$ is the ideal beamformer matrix. The $(r,s)$-th element of the multiplicative noise matrix is given by  $[\mathbf{E}_m]_{r,s} = e^{j \delta_{r,s}}$ with $ \delta_{r,s} \sim \mathcal{CN}(0, \epsilon_m^2)$, where $\epsilon_m$ controls the level of spillover loss. For the additive noise matrix we have that its entries are drawn independently as white Gaussian random variables, i.e., $[\mathbf{E}_a]_{r,s} \sim \mathcal{CN}(0, \epsilon_a^2)$.

Then, the received signal $\mathbf{r}$, measured at the outputs of the ADCs is expressed as
\begin{align}\label{eq:quantization_model}
\mathbf{r} = \mathcal{Q}(\mathbf{y}) &= \mathcal{Q}(\mathbf{F}_{\textrm{RF}}
\mathbf{H} \mathbf{s} +\mathbf{n}) \\
&= \mathcal{Q}((\mathbf{E}_m \circ \mathbf{F}_{\textrm{RF}}^{o})
\mathbf{H} \mathbf{s} + \mathbf{E}_a
\mathbf{H} \mathbf{s} + \mathbf{n})  \\
&= \mathcal{Q}(\underbrace{
(\mathbf{s}^T \otimes (\mathbf{E}_m \circ \mathbf{F}_{\textrm{RF}}^{o}))(\mathbf{D}_\textrm{T}^{*}\otimes \mathbf{D}_\textrm{R})}_{\mathbf{\Psi}} \underbrace{\textrm{vec}(\mathbf{Z})}_\mathbf{z} + \mathbf{d} + \mathbf{n}),
\end{align}
where $\mathbf{d} \in \mathbb{C}^{N_\textrm{R} \times 1}$ represents the beamforming noise, which is unknown in general since we have no knowledge of the noise matrix $\mathbf{E}_a$.

The unknown vector $\mathbf{d}$ is also a Gaussian random variable with covariance matrix given by:
\begin{equation}
\boldsymbol{\Sigma}_\textrm{d}^2 = \mathcal{E} \{ \mathbf{H}^H \mathbf{E}_a^H \mathbf{E}_a \mathbf{H} \} = \epsilon_a^2 \mathcal{E} \{\mathbf{H}^H \mathbf{H} \} = \epsilon_a^2 \mathcal{E}\{ \mathbf{Z}^H \mathbf{Z} \},
\end{equation}
given that $\mathcal{E}\{\Vert \mathbf{s} \Vert^2 \} =1$. Given that the channel has the line-of-sight (LoS) path, and very few weak non-LoS paths, the covariance matrix of the channel has strong diagonal structure, and thus it can be approximated by:
\begin{equation}
\mathcal{E}\{ \mathbf{Z}^H \mathbf{Z} \} \approx \sigma_\alpha^2 \mathbf{I},
\end{equation}
where $\sigma_alpha$ is the channel gain of the LoS path. Therefore, the covariance matrix of the noise vector $\mathbf{d}$ can be approximated by:
\begin{equation}
    \boldsymbol{\Sigma}^2_\textrm{d} \approx \sigma_\alpha^2 \epsilon_a^2 \mathbf{I}.
\end{equation}

The output can be expressed as the signal plus two noise vectors, i.e., $(\mathbf{d} + \mathbf{n})$. Finally the system model can be expressed as
\begin{align}
\mathbf{r} = \mathbf{\Psi}\mathbf{z} + (\mathbf{d} + \mathbf{n}),    
\end{align}
where the unknown channel $\mathbf{z}$ can be estimated using an efficient 
technique such as that shown in the following subsection.

It is well known that the quantization function $\mathcal{Q}(\cdot)$ is a non-linear function and therefore difficult to analyze directly. This makes the simple linear estimators such as the linear miminum mean square (LMMSE) inappropriate, resulting in high estimation errors. However, working on the statistical properties of the quantized signal allows strictly linear operations to be performed \cite{widrow}. To this end, we focus on the MAP estimator, which can be expressed as:
\begin{equation}
    \mathbf{\hat{h}} = \textrm{arg} \max_{\mathbf{z}, \mathbf{d}} p(\mathbf{r}, \mathbf{y}, \mathbf{d}, \mathbf{z}),
\end{equation}
with $p(\mathbf{r}, \mathbf{y},  \mathbf{d}, \mathbf{z})$ as the joint probability distribution function (PDF) of the quantized output, the received signal, the unknown noise and the unknown channel vector.

\subsection{Alternating Minimization}


In this subsection, we propose a MAP estimator to obtain the unknown channel vector $\mathbf{z}$. The proposed technique is based on the EM algorithm \cite{Gupta} which approximates the MAP solution. Specifically, EM is an iterative method that tries to find the maximum likelihood estimator of a parameter $\mathbf{z}$ of a parametric probability distribution $\mathbf{p(\mathbf{r} \vert \mathbf{z}})$. The quantities $\mathbf{y}$ and $\mathbf{d}$ are considered as unknown and they are computed by the EM approach \cite{mezghaniITG2010}.


\begin{proposition}
Given the noise covariance matrix $\mathbf{R}_\textrm{d}$, the following iterative EM algorithm converges to the optimum solution. Specifically, at the $(m+1)$-th iteration the following steps are performed:
\begin{itemize}
	\item \textbf{E-step} \\
	Compute the vector $\mathbf{b}^{(m+1)} \in \mathbb{R}^{2 L \times 1}$ with
		\begin{equation}\label{eq:estep_b}
		[\mathbf{b}^{(m+1)}]_i = - \frac{\sigma_\textrm{n}}{\sqrt{2 \pi}} \frac{  \xi(l_i) - \xi(u_i)}{\Xi(l_i) - \Xi(u_i) }
		\end{equation} 
		where $l_i, u_i$ are the lower/upper bounds of the quantizer for $[\mathbf{r}]_i$ respectively; 
		$$\xi(a) \triangleq e^{-\frac{(a - [\mathbf{\Psi}\mathbf{z}^{(m)}]_i - [\mathbf{d}]_i)^2}{2 \sigma^2}},$$
		$$\Xi(a) \triangleq \mathrm{erf}(\frac{-a + [\mathbf{\Psi}\mathbf{z}^{(m)}]_i + [\mathbf{d}]_i}{\sqrt{2} \sigma_\textrm{n}}),$$
		where $\mathrm{erf}(\cdot)$ is the error function.
	\item \textbf{M-step}
	\begin{itemize}
	    \item Estimate the channel vector $\mathbf{z}^{(m+1)} \in \mathbb{R}^{2 L \times 1}$ by estimating the linear system of equations:
		\begin{equation}\label{eq:mstep_system}
		\mathbf{A} \mathbf{z}^{(m+1)} = \mathbf{\Psi}^T \mathbf{c},
		\end{equation} 
		where
		$$\mathbf{A} \triangleq \mathbf{\Psi}^T \mathbf{\Psi} + \sigma_\textrm{n}^2 \mathbf{I}_\textrm{L},$$
		$$\mathbf{c} \triangleq \mathbf{\Psi} \mathbf{z}^{(m)} + \mathbf{b}^{(m+1)} + \mathbf{d}^{(m)}$$
		\item Obtain the update of the noise vector:
		\begin{equation}\label{eq:update_d}
		    \mathbf{d}^{(m+1)} = \mathbf{R}_\textrm{d} \mathbf{\Psi} \mathbf{z}^{(m+1)}
		\end{equation}
	\end{itemize}
\end{itemize}
\end{proposition}
\subsubsection*{Proof} The joint PDF can be written as: $
p(\mathbf{r}, \mathbf{y}, \mathbf{d}, \mathbf{z}) = p(\mathbf{y},\mathbf{d} \vert \mathbf{r},\mathbf{z}) p(\mathbf{r}\vert \mathbf{y}) p(\mathbf{z})
= p(\mathbf{y} \vert\mathbf{d}, \mathbf{r},\mathbf{z}) p(\mathbf{r}\vert \mathbf{y}) p(\mathbf{d}) p(\mathbf{z})$. Hence, taking the expectation over the unknown variables we have:
\begin{align}
     & \mathcal{E}_{\mathbf{y}, \mathbf{d} \vert \mathbf{r}, \mathbf{z}^{(m)}}\left\{ \ln p(\mathbf{r}, \mathbf{y}, \mathbf{d}, \mathbf{z}) \right\} = \mathcal{E}_{\mathbf{y}, \mathbf{d} \vert \mathbf{r}, \mathbf{z}^{(m)}}\left\{ \ln p(\mathbf{y},\mathbf{r} \vert \mathbf{d}, \mathbf{z}) \right\} \nonumber \\ & \hspace{3em} + \ln p(\mathbf{r} \vert \mathbf{y}) + \mathcal{E}_{\mathbf{d} \vert \mathbf{r}, \mathbf{z}^{(m)}}\left\{ \ln p(\mathbf{d})\right\} + \ln p(\mathbf{z})
\end{align}
The first term can be written as:
$\mathcal{E}_{\mathbf{y}, \mathbf{d} \vert \mathbf{r}, \mathbf{z}^{(m)}}\left\{ \ln p(\mathbf{y} \vert \mathbf{r}, \mathbf{d}, \mathbf{z}) \right\} = - \frac{1}{2 \sigma_\textrm{n}^2} \mathcal{E}\left\{ \Vert \mathbf{y} - \mathbf{\Psi} \mathbf{z} - \mathbf{d} \Vert_2^2 \right\} + \kappa_1$, with $\kappa_1$ a constant term. The expectation term is further expanded as:
\begin{align}
    & \mathcal{E}_{\mathbf{y}, \mathbf{d} \vert \mathbf{r}, \mathbf{z}^{(m)}}\left\{ \Vert \mathbf{y} - \mathbf{\Psi} \mathbf{z} - \mathbf{d} \Vert_2^2 \right\}  =  \mathcal{E}_{\mathbf{y}, \mathbf{d} \vert \mathbf{r}, \mathbf{z}^{(m)}} \left\{ \Vert \mathbf{y} - \mathbf{d} \Vert_2^2 \right\} \nonumber \\ & \hspace{4em}
    - 2\mathcal{E}_{\mathbf{y}, \mathbf{d} \vert \mathbf{r}, \mathbf{z}^{(m)}} \left\{ (\mathbf{y} - \mathbf{d})^T \mathbf{\Psi} \mathbf{z}\right\} + \Vert \mathbf{\Psi} \mathbf{z}\Vert_2^2.
\end{align}
Keeping only the terms that depend on $\mathbf{z}$ and $\mathbf{d}$ we define the function:
\begin{align}
\mathcal{G} &\triangleq -\frac{1}{2 \sigma_\textrm{n}^2} \left(- 2\mathcal{E}_{\mathbf{y}, \mathbf{d} \vert \mathbf{r}, \mathbf{z}^{(m)}} \left\{ (\mathbf{y} -\mathbf{d})^T \mathbf{\Psi} \mathbf{z}\right\} + \Vert \mathbf{\Psi} \mathbf{z}\Vert_2^2 \right. \nonumber \\
    &\left. +\mathcal{E}_{\mathbf{y}, \mathbf{d} \vert \mathbf{r}, \mathbf{z}^{(m)}} \left\{ - 2\mathbf{y}^T \mathbf{d} + \Vert \mathbf{d} \Vert_2^2\right\}  \right) + \ln p(\mathbf{z})  \\
    &=   \frac{1}{2 \sigma_\textrm{n}^2} \left(2\mathcal{E}_{\mathbf{y}, \mathbf{d} \vert \mathbf{r}, \mathbf{z}^{(m)}} \left\{ \mathbf{y}\right\}^T \mathbf{\Psi} \mathbf{z} \label{eq:tranc_normal} + 2 \mathbf{d}^T \mathbf{\Psi} \mathbf{z} - \Vert \mathbf{\Psi} \mathbf{z}\Vert_2^2\right.  \nonumber \\
    & \left. - \Vert \mathbf{d} \Vert_2^2 + 2\mathcal{E}_{\mathbf{y}, \mathbf{d} \vert \mathbf{r}, \mathbf{z}^{(m)}} \left\{ \mathbf{y}\right\}^T \mathbf{d} \right) - \frac{1}{2 \sigma_\textrm{s}^2} \Vert \mathbf{z} \Vert_2^2 
\end{align}
where \eqref{eq:tranc_normal} represents the mean of the truncated Gaussian random variable. Hence,
\begin{equation}
    \mathcal{E}_{\mathbf{y}, \mathbf{d} \vert \mathbf{r}, \mathbf{z}^{(m)}} \left\{ [\mathbf{y}]_i\right\} = [\mathbf{\Psi} \mathbf{z}^{(m)}]_i + \sigma_n \frac{\xi(r_i^{\textrm{lo}}) -\xi(r_i^{\textrm{up}})}{\Xi(r_i^{\textrm{up}}) - \Xi(r_i^{\textrm{lo}})}.
\end{equation}

Taking the derivative of $\mathcal{G}$ over $\mathbf{z}$ and setting it equal to zero, we have:
\begin{align}
    \frac{\partial \mathcal{G}}{\partial \mathbf{z}} = 0 
    \Rightarrow 2 \mathbf{\Psi}^T \mathbf{c} - 2 (\mathbf{\Psi}^T \mathbf{\Psi} + \frac{\sigma_\textrm{n}^2}{\sigma_\textrm{s}^2}) \mathbf{z}  =0
\end{align}
where $\mathbf{c} \triangleq [\mathbf{\Psi} \mathbf{z}^{(m)}]_i + \sigma_\textrm{n} \frac{\xi(r_i^{\textrm{lo}}) -\xi(r_i^{\textrm{up}})}{\Xi(r_i^{\textrm{up}}) - \Xi(r_i^{\textrm{lo}})} - \mathbf{d}$.

Taking the derivative of $\mathcal{G}$ over $\mathbf{d}$ and setting it equal to zero, we have:
\begin{align}
    \frac{\partial \mathcal{G}}{\partial \mathbf{d}} = 0
    \Rightarrow  \mathbf{d} = \mathbf{R}_\textrm{d} \mathbf{\Psi} \mathbf{z}.
\end{align}

\subsubsection*{Complexity} The computational complexity order of the EM algorithm is mainly determined by the complexity of equation \eqref{eq:mstep_system}. The complexity order of \eqref{eq:estep_b} and \eqref{eq:update_d} is only $\mathcal{O}(M)$ while the number of the iterations which are required for convergence is usually very small (e.g., 10-20).

\section{Simulation Results}
In this section, we evaluate the performance of the proposed EM-based technique via computer simulations. The results have been averaged over $100$ Monte-Carlo realizations. 

\begin{figure}[t]
    \vspace{1em}
	\centering
	\includegraphics[width=0.5\textwidth]{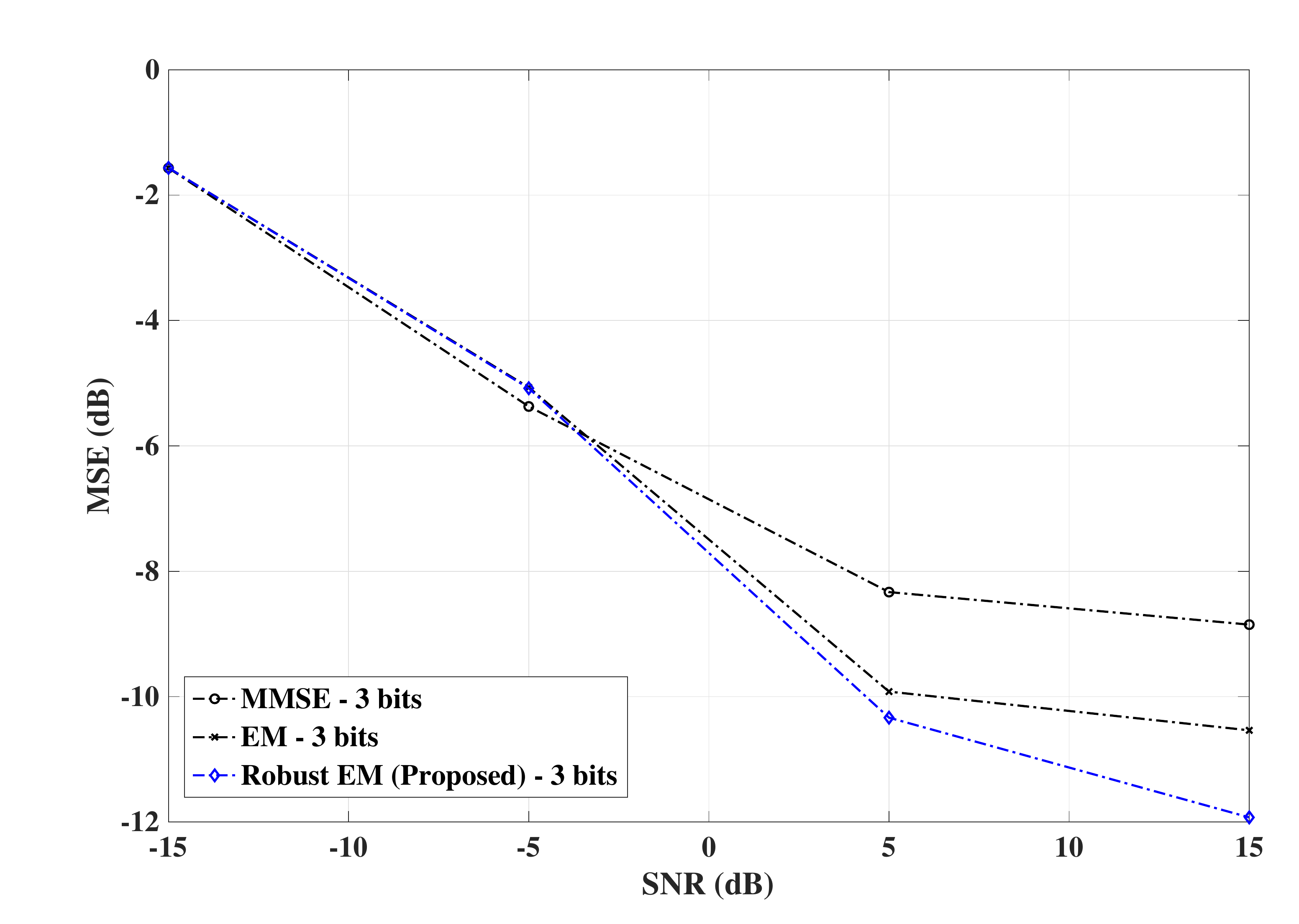}
	\caption{MSE versus SNR with 3-bit, $T=200$, $N_\textrm{R}=15$, $L = 9$ and $\sigma_\textrm{d}^2=0.012$.}
	\label{mse_vs_snr}
\end{figure}

\begin{figure}[t]
    \vspace{1em}
	\centering
	\includegraphics[width=0.5\textwidth]{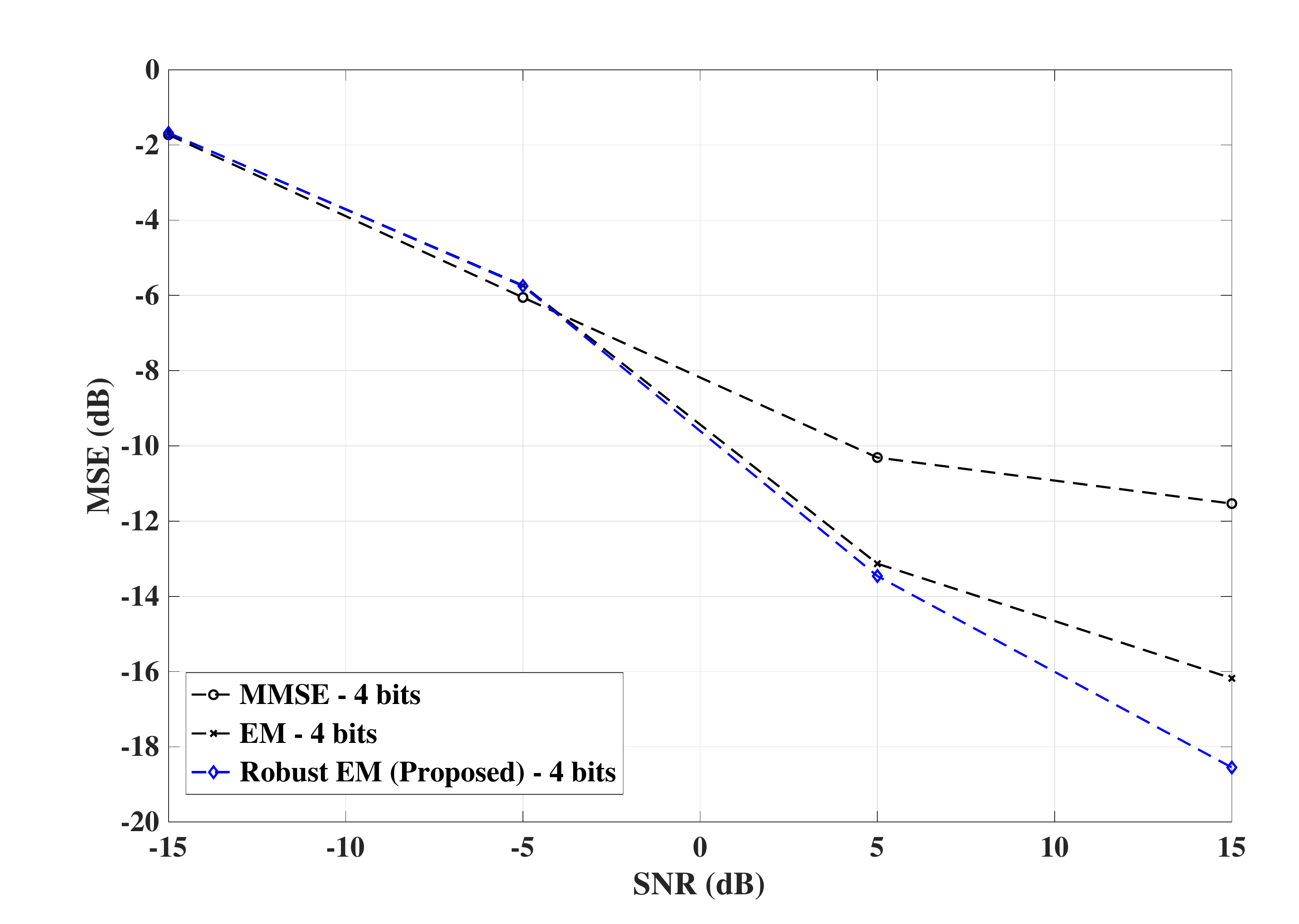}
	\caption{MSE versus SNR with 4-bit, $T=200$, $N_\textrm{R}=15$, $L = 9$ and $\sigma_\textrm{d}^2=0.012$.}
	\label{mse_vs_snr}
\end{figure}

\begin{figure}[t]
    \vspace{1em}
	\centering
	\includegraphics[width=0.5\textwidth]{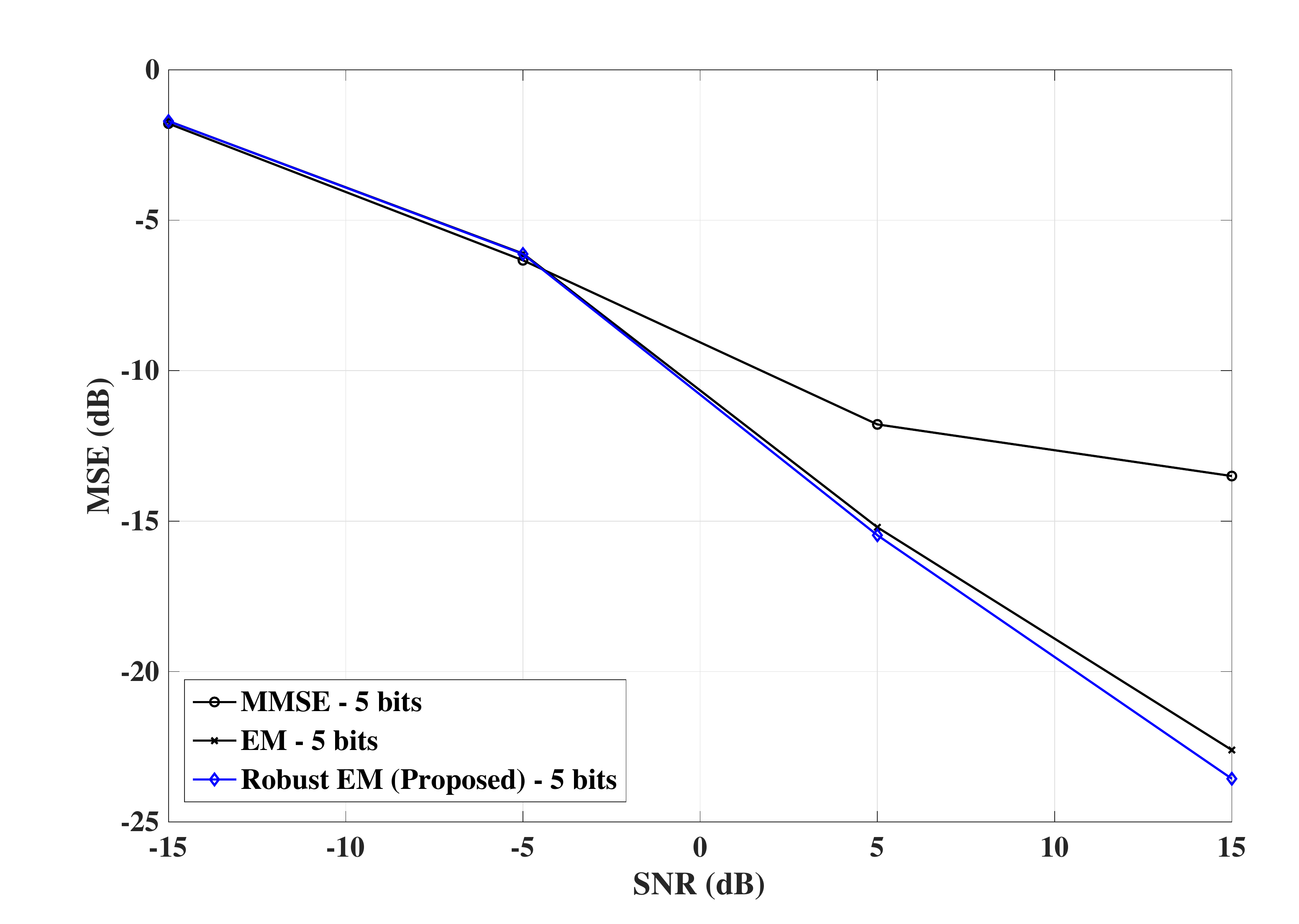}
	\caption{MSE versus SNR with 5-bit, $T=200$, $N_\textrm{R}=15$, $L = 9$ and $\sigma_\textrm{d}^2=0.012$.}
	\label{mse_vs_snr}
\end{figure}

\subsubsection*{System Setup}
The mmWave channel is simulated at a frequency of 28 GHz assuming 4 scattering clusters with a total number of 5 sub-paths for each cluster, i.e., 20 multipaths in total. The instantaneous path gains are modeled based on Gaussian distribution with zero-mean and unit variance. The UE transmits $L$ Gaussian symbols with repetition coding of length $T$. The average transmit power is set to $\rho^{1/2}=0.1$. We consider the range of 3-5 bit resolutions for the ADCs.

We use the classical Rotman lens principles to design all the lenses. The first stage lens, i.e., right after the URA, consists of $N_{\textrm{ap}} = 5$ and $N_{\textrm{bp}} = 3$, followed by the second stage lens having $N_{\textrm{ap}} = N_{\textrm{bp}} = 3$. We use number of antennas in this topology as $N_\textrm{R} = 15$ and the number of RF chains as $L = 9$. 
The first stage lens is implemented for beamfoming along azimuth, while the second stage lenses handle beamforming in elevation. URA antenna elements are operational at 28 GHz.


\subsubsection*{Estimation Performance}
We compare the performance of the proposed EM-based channel estimator with the conventional EM estimator \cite{Gupta} and the MMSE approach. The results for the proposed technique and the baseline approaches are shown in Figs. 1-3 for the cases of 3, 4 and 5 bit ADC resolutions, respectively. We can observe that proposed EM-based approach outperforms the conventional EM approach and the MMSE approach for all the different bit cases. For example, at $5$ dB SNR and $3$-bit case, it can be observed that proposed robust EM-based method performs $2.25$ dB better and $0.25$ dB better than the SNR performance of the conventional EM approach and the MMSE approach, respectively. In another example, at $5$ dB SNR and for the $5$-bit case, the proposed approach is around $3.5$ dB better and around $0.75$ dB better than the SNR performance of the conventional EM approach and the MMSE approach, respectively.

\section{Conclusions}
This paper considers a mmWave hybrid MIMO RX with a Rotman lens architecture antenna array and low-resolution ADCs. 
This design provides a trade-off between the hardware complexity and the introduced noise to the received signal. We investigate the performance of the lens-based system with low resolution ADCs in terms of the mmWave channel estimation. We develop a robust MAP estimator based on EM iterative algorithm, which outperforms the conventional EM approach in terms of MSE. This is achieved without increasing the hardware complexity of the system, while the baseband signal processing complexity increases only slightly.

\end{document}